\documentclass[twocolumn]{aastex6}

\shorttitle{Kepler Habitable Zone Exoplanet Candidates}
\shortauthors{Stephen R. Kane et al.}
\slugcomment{Submitted for publication in the Astrophysical Journal}

\begin{document}

\title{A Catalog of Kepler Habitable Zone Exoplanet Candidates}

\author{
  Stephen R. Kane\altaffilmark{1},
  Michelle L. Hill\altaffilmark{1},
  James F. Kasting\altaffilmark{2},
  Ravi Kumar Kopparapu\altaffilmark{3},
  Elisa V. Quintana\altaffilmark{4},
  Thomas Barclay\altaffilmark{4},
  Natalie M. Batalha\altaffilmark{4},
  William J. Borucki\altaffilmark{4},
  David R. Ciardi\altaffilmark{5},
  Nader Haghighipour\altaffilmark{6},
  Natalie R. Hinkel\altaffilmark{1,7},
  Lisa Kaltenegger\altaffilmark{8},
  Franck Selsis\altaffilmark{9},
  Guillermo Torres\altaffilmark{10}
}
\email{skane@sfsu.edu}
\altaffiltext{1}{Department of Physics \& Astronomy, San Francisco
  State University, 1600 Holloway Avenue, San Francisco, CA 94132,
  USA}
\altaffiltext{2}{Department of Geosciences, Penn State University, 443
  Deike Building, University Park, PA 16802, USA}
\altaffiltext{3}{NASA Goddard Space Flight Center, 8800 Greenbelt
  Road, Mail Stop 699.0 Building 34, Greenbelt, MD 20771, USA}
\altaffiltext{4}{NASA Ames Research Center, Moffett Field, CA 94035,
  USA}
\altaffiltext{5}{NASA Exoplanet Science Institute, Caltech, MS 100-22,
  770 South Wilson Avenue, Pasadena, CA 91125, USA}
\altaffiltext{6}{University of Hawaii-Manoa, Honolulu, HI 96822, USA}
\altaffiltext{7}{School of Earth \& Space Exploration, Arizona State
  University, Tempe, AZ 85287, USA}
\altaffiltext{8}{Carl Sagan Institute, Cornell University, Ithaca, NY
  14853, USA}
\altaffiltext{9}{Laboratoire d'astrophysique de Bordeaux,
  Univ. Bordeaux, CNRS, B18N, all\'ee Geoffroy Saint-Hilaire, 33615
  Pessac, France}
\altaffiltext{10}{Harvard-Smithsonian Center for Astrophysics, 60
  Garden Street, Cambridge, MA 02138, USA}


\begin{abstract}

The NASA {\it Kepler} mission has discovered thousands of new
planetary candidates, many of which have been confirmed through
follow-up observations. A primary goal of the mission is to determine
the occurrance rate of terrestrial-size planets within the Habitable
Zone (HZ) of their host stars. Here we provide a list of HZ exoplanet
candidates from the {\it Kepler} Data Release 24 Q1-Q17 data vetting
process. This work was undertaken as part of the {\it Kepler}
Habitable Zone Working Group. We use a variety of criteria regarding
HZ boundaries and planetary sizes to produce complete lists of HZ
candidates, including a catalog of 104 candidates within the
optimistic HZ and 20 candidates with radii less than two Earth radii
within the conservative HZ. We cross-match our HZ candidates with the
Data Release 25 stellar properties and confirmed planet properties to
provide robust stellar parameters and candidate dispositions. We also
include false positive probabilities recently calculated by
\citet{mor16} for each of the candidates within our catalogs to aid in
their validation. Finally, we performed dynamical analysis simulations
for multi-planet systems that contain candidates with radii less than
two Earth radii as a step toward validation of those systems.

\end{abstract}



\section{Introduction}
\label{introduction}

The last few decades have seen an extraordinary progression in the
field of exoplanetary science. The rate of exoplanet discovery has
continued to increase as the sensitivity to smaller planets has
dramatically improved. The discoveries of the {\it Kepler} mission
have had a major impact on our understanding of exoplanet orbit, size,
and multiplicity distributions \citep{lis14,row14}. The primary source
of {\it Kepler} discoveries to the scientific community has been
through the regular release and update of exoplanetary candidates
\citep{bor11a,bor11b,bat13,bur14,row15,mul15,cou16}. These discoveries
have shown that the frequency of planets increases to smaller sizes;
thus terrestrial planets are more common than gas giant planets
\citep{fre13,how13,pet13}.

The significance of a terrestrial-planet-rich universe is fully
realized in the study of habitability. The {\it Kepler} mission
\citep{bor16} has a primary science goal of determining the frequency
of terrestrial planets in the Habitable Zone (HZ): usually defined as
the region around a star where water can exist in a liquid state on
the surface of a planet with sufficient atmospheric pressure
\citep{kas93}. Commonly referred to as eta-Earth ($\eta_{\oplus}$),
the frequency of HZ terrestrial planets has become a major focus of
interpreting {\it Kepler} results
\citep{cat11,tra12,dre13,gai13,kop13b,for14,mor14,dre15}. The process
of determining eta-Earth requires a reliable list of HZ candidates
whose properties have been adequately vetted to produce robust
planetary and stellar properties.

Here we present an exhaustive catalog of HZ candidates from the Q1-Q17
Data Release 24 (DR24) candidate list, along with an analysis of the
radii distributions and orbital stabilities. The work described here
is the product of efforts undertaken by the {\it Kepler} Habitable
Zone Working Group. The Q1-Q17 DR24 catalog heavily favors uniform
vetting over the correct dispositions of individual objects, in order
to be principally used to calculate statistically accurate occurrence
rates. We cross-match the DR24 candidates with both revised stellar
parameters and confirmed planets to provide the most complete list of
HZ candidates from the {\it Kepler} mission. In Section \ref{hzb} we
describe the adopted boundaries for the HZ. Section \ref{hzcand}
presents four different HZ criteria for which we present tables and
statistics for candidates in each category and examine their
distributions. The stability of HZ planet candidates in multi-planet
systems is a necessary step in fully characterizing such planets, and
we provide the results of such analyses in
Section~\ref{stability}. Finally, we provide concluding remarks and
proposals for further work in Section \ref{conclusions}.


\section{Habitable Zone Boundaries}
\label{hzb}

The {\it Kepler} mission has provided several cases of confirmed
planets of terrestrial size that lie in the HZ of their host star.
\citep{bor12,qui14,tor15}. The concept of the HZ has appeared in the
literature for some time \citep{hua59,hua60,har78,har79}, but only in
recent decades have complex climate models been brought to bear on the
problem of quantifying the boundaries. A key conceptual development
was the inclusion of CO$_2$-climate feedback, introduced by
\citet{kas93}. (Note that this feedback was also included in the
\citet{har78,har79} models, but the greenhouse effect of CO$_2$ was
underestimated and thus he concluded that frozen planets could never
deglaciate.) Importantly for our purposes, the \citet{kas93} model
included three sets of possible boundaries. On the inner edge, these
were the moist greenhouse (in which water started to be lost), the
runaway greenhouse (in which the oceans evaporate entirely), and the
``recent Venus'' limit (based on the empirical observation that the
surface of Venus has been dry for at least a billion years). On the
outer edge, the proposed limits were the ``1st CO$_2$ condensation''
limit (where CO$_2$ condensation first occurs), the maximum greenhouse
limit (where the CO$_2$ greenhouse effect maximizes), and the ``early
Mars'' limit (based on the observation that Mars appears to have been
habitable 3.8~Gyrs ago when solar luminosity was some 25\% lower.

\begin{figure}
  \includegraphics[width=9.0cm]{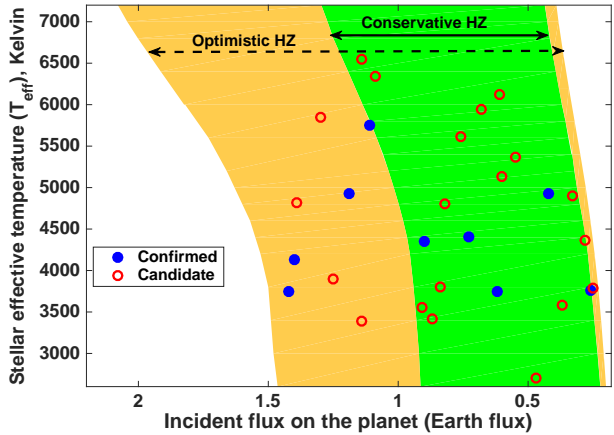}
  \caption{Stellar effective temperature as a function of incident
    flux for the unconfirmed candidates (open red circles) and
    confirmed planets (solid blue circles) from
    Table~\ref{c2cand}. These are overplotted on the conservative and
    optimistic HZ.}
  \label{hzfig}
\end{figure}

Since that time, these 1-D habitability limits have been re-evaluated
using updated absorption coefficients for CO$_2$ and H$_2$O
\citep{kop13a,kop14}. With the new coefficients, the moist and runaway
greenhouse limits on the inner edge have coalesced. The ``1st
condensation'' limit on the outer edge was abandoned well before this,
because calculations suggested that CO$_2$ clouds should generally
warm the climate rather than cool it \citep{for97,mis00}. This
conclusion has since been revised. The early CO$_2$ cloud studies used
two-stream approximation \citep{too89} for radiative transfer -- a
method that evidently overestimates the transmitted and reflected
radiation, thereby yielding a higher scattering greenhouse effect
\citep{kit13}. When \citet{kit13} used a higher-order discrete
ordinate method, DISORT, with 24 radiation streams, they found a much
smaller warming. Nevertheless, CO$_2$ clouds still do not cool a
planet strongly, and so the 1st condensation limit on the outer edge
can still be ignored. Now it is often considered that there are two
limits at each HZ boundary, one theoretical and one empirical. The two
limits for the outer edge are nearly the same, about 1.7--1.8~AU for a
Sun-like star. At the inner edge, though, the theoretical runaway
greenhouse limit from the \citet{kop14} model is 0.99~AU, whereas the
recent Venus limit remains at 0.75~AU (Venus itself is at
0.72~AU). The solar flux difference between the empirical and
theoretical inner edges is a factor of $(0.99/0.75)^2 \cong 1.7$;
thus, it makes sense to talk about a ``conservative'' HZ
(0.99--1.7~AU) and an ``optimistic'' HZ (0.75--1.8~AU). Note that, as
described below, the inner edge calculated by 1-D models is almost
certainly overly conservative, and 0.95~AU is a better estimate for
the inner HZ boundary. These limits are shown in Figure~\ref{hzfig} as
a function of the flux from the star normalized to the flux at Earth's
orbit. The boundaries vary for different stellar types because of the
different albedo of an Earth-like planet under different wavelengths
of stellar irradiation. HZ calculations for known exoplanet systems
using these conservative and optimistic limits are available through
the Habitable Zone Gallery\footnote{\tt http://hzgallery.org}
\citep{kan12a} and described in more detail by \cite{kan13}. A HZ
calculator is also available via the website of the Virtual Planetary
Laboratory\footnote{\tt
  http://depts.washington.edu/naivpl/content/hz-calculator}.

Determining which of the HZ definitions, conservative or optimistic,
is more useful depends on the task at hand. \citet{kas14} have argued
that a conservative definition should be adopted for purposes of
calculating eta-Earth. That is because this parameter may eventually
be used to estimate the size of a future flagship telescope mission
designed to find and characterize such planets. But once such a
telescope is launched and returning data, a more optimistic definition
may need to be adopted in order to avoid inadvertently neglecting
exoplanets that lie within the broader, empirical HZ.

Some authors have proposed modifications to the HZ limits based on
additional greenhouse gases (see \citet{sea13} for
review). Specifically, accumulation of significant amounts of
molecular hydrogen (H$_2$) in a planet's atmosphere can extend the
outer edge of the HZ dramatically \citep{ste99,pie11}. Molecular
hydrogen condenses only at very low temperatures, and its
collision-induced absorption encompasses the entire thermal-infrared
spectrum \citep{wor13}. \citet{pie11} showed that a 3 Earth-mass
planet with 40-bar H$_2$ atmosphere can maintain a surface temperature
of 280~K at 10~AU from a G-type star. Even free-floating terrestrial
planets, with dense enough H$_2$ atmospheres, could remain habitable
provided that they had sufficient internal heat \citep{ste99}. But
while such far flung planets may exist, it is not clear that we should
allow them to influence the design of a future direct imaging
telescope to observe potential habitable planets. The contrast ratio
between the Earth and the Sun is $\sim 10^{-10}$ in the visible
\citep{lev06}, so an Earth-like planet at 10~AU, with a similar
albedo, would have a contrast ratio 100 times smaller, making it
difficult to observe. And free-floating habitable planets, which have
an effective radiating temperature of $\sim$30~K \citep{ste99}, would
be virtually impossible to detect remotely. So, it is better to accept
a conservatively defined HZ for now, bearing in mind that some planets
beyond this range might still be habitable.

It should also be recognized that the theoretical HZ limits are
evolving with time as climate models improve. 3-D climate models can
include factors such as relative humidity variations and clouds that
are impossible to estimate accurately in 1-D calculations. A recent
3-D study by \citet{lec13} shows that the inner HZ edge for a Sun-like
star moves in to at least 0.95~AU because of low relative humidity in
the descending branches of the tropical Hadley cells, convection cells
in which air rises at the equator and sinks at medium
latitudes. Another study by \citet{wol14} suggests that the inner edge
can be even closer, at 0.93~AU, because of negative feedback from
clouds. And \citet{yan13,yan14} have argued that the inner HZ edge for
synchronously rotating planets around late-K and M stars could occur
at a stellar flux equal to twice that at Earth's orbit because of
widespread cloudiness on their sunlight sides. \citet{kop16} noted
that, correcting \citet{yan13,yan14} studies with consistent orbital
periods, the inner edge of the HZ around M-dwarfs is further away than
proposed by those studies. Nevertheless, \citet{kop16} confirmed the
substellar cloud mechanism originally proposed by \citet{yan13}. A
recent calculation by \citet{lec15} shows that atmospheric thermal
tides on M-star planets can prevent synchronous planetary
rotation. Such an effect can potentially jeopardize habitability if
synchronization is required to ensure a sufficient albedo, but can
also favor habitability by increasing heat redistribution efficiency.

A related issue concerning the inner edge of the HZ has to do with dry
planets, sometimes called ``Dune'' planets after the science fiction
novel by that name. A (low-obliquity) Dune planet would be mostly
desert but would have water-rich oases near its poles. Such planets
can, in theory, remain habitable closer in to its parent star because
the positive feedback caused by water vapor would be much weaker in
this case. \citet{abe11} simulated such a planet with a highly
parameterized 3-D climate model and determined that the inner HZ edge
for a Sun-like star could be as close as 0.77~AU, near the empirical
``recent Venus'' limit. More recently, \citet{zso13} approached the
same problem with a 1-D climate model and determined an inner edge of
0.38~AU. However, this result has been criticized by \citet{kas14},
who argue that a 1-D model is not appropriate for such an inherently
3-D problem, as it does not explicitly identify regions where surface
liquid water might exist.

As mentioned earlier, we suggest using conservative estimates of the
HZ for planet occurrence rate studies from \citet{kop14}.  Some 3-D
climate modeling studies have been used to estimate the HZ limits, but
it may take time for different models to reach consensus. For this
reason, in this study, we provide candidates that lie within both the
conservative and optimistic estimates of the HZ from the 1-D model
study of \citet{kop14}, which encompass the limits from 3-D
models.


\section{Kepler Habitable Zone Candidates}
\label{hzcand}

We extracted the {\it Kepler} candidates from the NASA Exoplanet
Archive\footnote{\tt http://exoplanetarchive.ipac.caltech.edu}
\citep{ake13}. The associated data are current as of April 26, 2016
and are derived from the Data Release 24 (DR24) table of candidates
\citep{cou16}. In order to perform the necessary calculations, we
required the stellar effective temperature ($T_\mathrm{eff}$) and
radius ($R_\star$), as well as the planetary parameters of semi-major
axis ($a$) and radius ($R_p$). We utilize the revised stellar
parameters from the Data Release 25 (DR25) stellar table
\citep{mat16,twi16} to obtain $T_\mathrm{eff}$ and $R_\star$ values,
and recalculate $R_p$ and its uncertainty using the $R_p/R_\star$
values from DR24 and the $R\star$ values from DR25. Similarly, the
semi-major axes are recalculated using the DR25 $M_\star$ values for
self-consistency. The HZ boundaries were calculated using the
methodology described in Section~\ref{hzb} and by \citet{kan12a}. Note
that the reason cross-matching the DR24 and DR25 tables is necessary
is because, although the DR25 is more recent, it only contains stellar
information for the candidates. Also note that the planetary radius is
not needed for the HZ calculations, but is required for the
categorization process described below. We also calculate the incident
flux received by the planet ($F_p$) in units of the solar constant:
\begin{equation}
  \frac{F_p}{F_\oplus} = \left( \frac{R_\star}{R_\odot} \right)^2
  \left( \frac{T_\mathrm{eff}}{T_{\mathrm{eff},\odot}} \right)^4
  \left( \frac{\mathrm{1 AU}}{a} \right)^2
\end{equation}
The number of candidates for which we were able to extract all of the
needed information for our calculations was 4,270. Those candidates
for which there was insufficient information were noted by
\citep{cou16} as likely having very low signal-to-noise and are low
confidence candidates.

We separate the {\it Kepler} candidates into four categories. Category
1 candidates are in the conservative HZ and have a radius of $R_p < 2
R_\oplus$. Category 2 candidates are in the optimistic HZ and have a
radius of $R_p < 2 R_\oplus$. Category 3 candidates are in the
conservative HZ and can have any radius. Category 4 candidates are in
the optimistic HZ and can have any radius. Clearly this means that
some categories are subsets of others. For example, category 1 is a
subset of category 2, and category 4 contains all of the candidates
from categories 1--3. The number of exoplanet candidates in each
category are 20, 29, 63, and 104 for categories 1, 2, 3, and 4
respectively. The identifiers and relevant stellar and planet
parameters are shown in Tables \ref{c1cand}, \ref{c2cand},
\ref{c3cand}, and \ref{c4cand}. A handful of cases have parameter
uncertainties of zero due to incomplete information in the {\it
  Kepler} data release. In cases where the candidate has been
confirmed in the literature, we include the Kepler name in the second
column and replace the planetary and stellar parameters with those
from the relevant publications. The specific cases are Kepler-22~b
\citep{bor12}; Kepler-62~e \& f \citep{bor13}; Kepler-174~d,
Kepler-283~c, Kepler-298~d, Kepler-309~c, Kepler-315~c \citep{row14};
Kepler-186~f, Kepler-440~b, Kepler-442~b, Kepler-443~b \citep{tor15};
Kepler-296~e \& f \citep{bar15}; and Kepler-452~b \citep{jen15}. The
Table~\ref{c2cand} candidates (open red circles) and confirmed planets
(solid blue circles) are plotted with respect to the conservative and
optimistic HZ regions in Figure~\ref{hzfig}.

Note that, even though there is a broad consensus that the boundary
between terrestrial and gaseous planets likely lies close to $1.6
R_\oplus$ \citep{wei14,rog15,wol15}, we use $2 R_\oplus$ as our cutoff
to account for uncertainties in the stellar and planetary parameters
that would remove potentially terrestrial planets from our 1 \& 2
category lists. Such a safeguard is particularly relevant in light of
the fact that blended binaries may cause many of the candidate radii
to be underestimated \citep{car15,cia15,gil15}. Recent observations of
{\it Kepler} candidates by \citet{kra16} revealed wide binary
companions to the following candidates from
Tables~\ref{c1cand}--\ref{c4cand}: K00087, K00571, K00701, K00854,
K01298 K01422, K01431, K02418, K02626, K02686, K02992, K03263,
K04016. The presence of a previously undetected companion can affect
HZ boundaries within the system due to the additional source of
stellar flux \citep{kal13}, and also may impact the derived depth, and
thus radius, of a planet candidate, if the flux from the newly
detected companion is a significant fraction of the host star.

\begin{figure}
  \includegraphics[width=8.2cm]{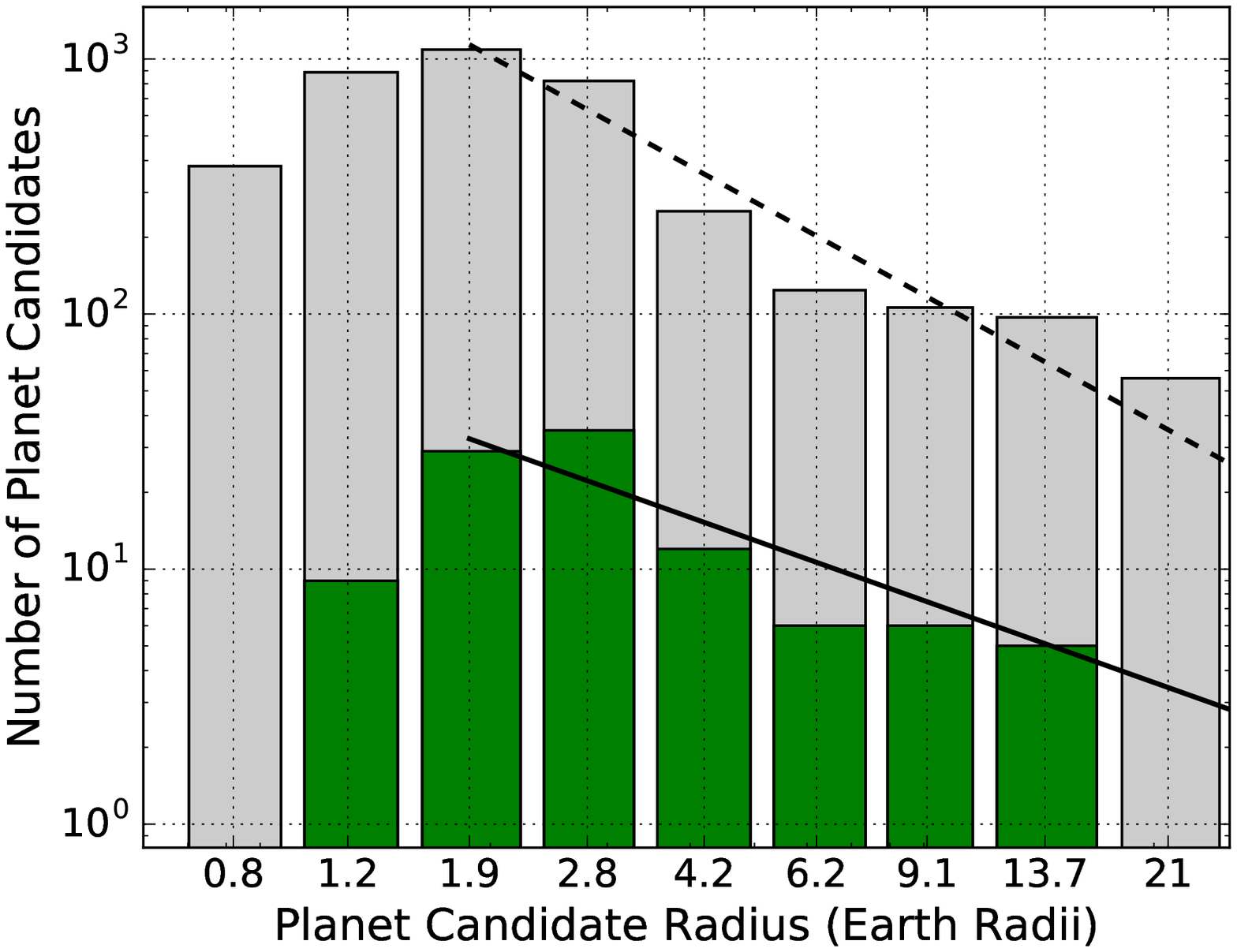} \\
  \includegraphics[width=8.2cm]{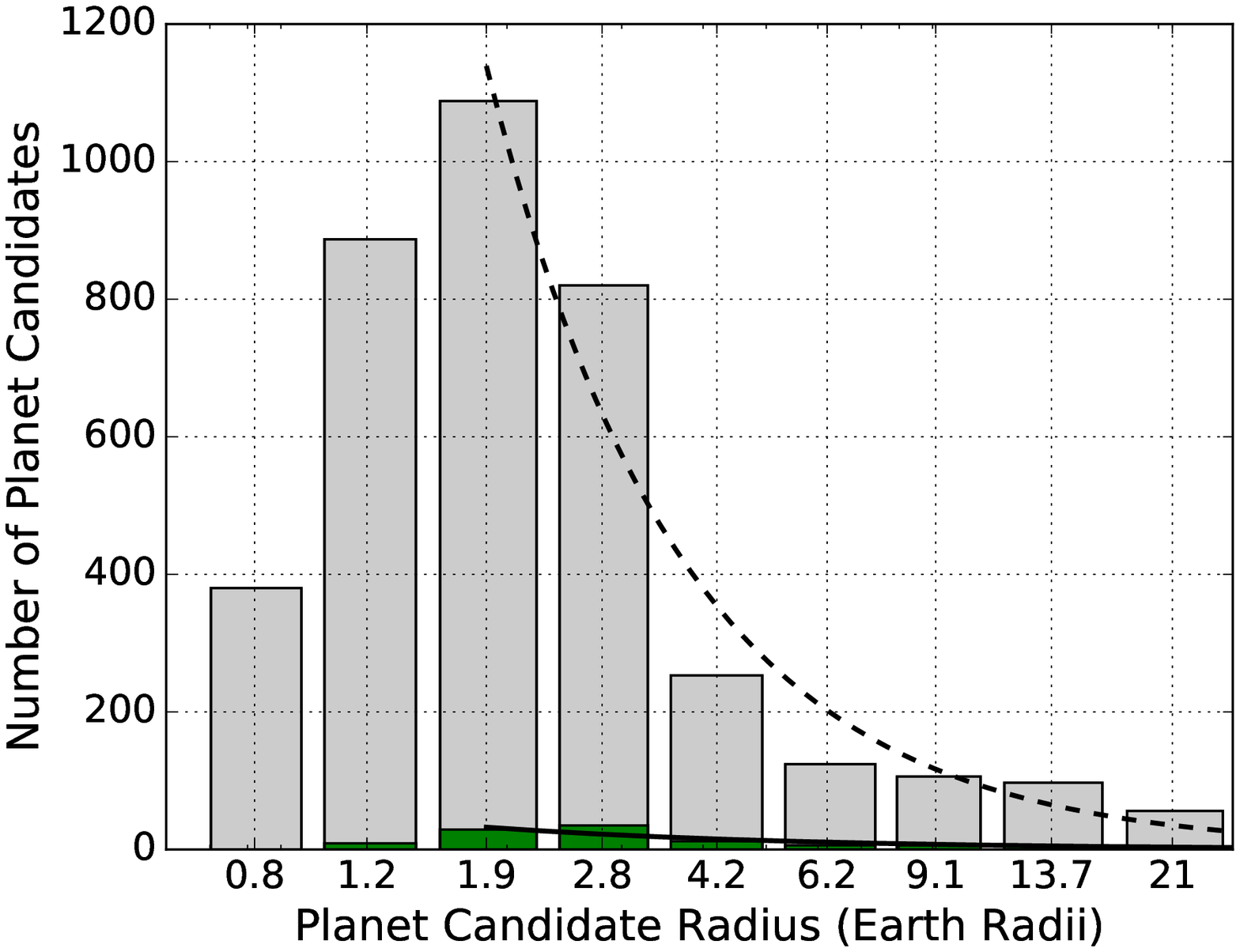}
  \caption{Histograms of all {\it Kepler} candidate radii (gray)
    relative to those candidates that are in the optimistic HZ of
    their host star (green), shown on a logarithmic (top panel) and
    linear (bottom panel) scale. The solid lines are power law fits to
    the HZ candidates and the dashed lines are power law fits to the
    entire {\it Kepler} distribution. Statistical analysis of the
    distributions shows that there is little evidence of a significant
    difference in the populations.}
  \label{histograms}
\end{figure}

Determining the false positive rate (FPR) of {\it Kepler} candidates
has been an on-going area of analysis for the past several years
\citep{mor11,san12,fre13}. Results from an analysis by \citet{des15}
indicate a relatively low FPR of 8.8\% for {\it Kepler} candidates
based on {\it Spitzer} observations and other follow-up data. For
candidates with periods larger than 9--12 months, additional care must
be taken to avoid false positives due to spacecraft-related systematic
noise, such as artifacts on specific detector channels
\citep{ten13,tor15}. Manual inspection by \citet{cou16} found that
following candidates from Tables~\ref{c1cand}--\ref{c4cand} are likely
instrumental artifacts: K06343.01, K06425.01, K07235.01, K07470.01,
K07554.01, K07591.01. Recent work by \citet{mor16} uses an automated
probabilistic validation to produce false positive probabilities (FPP)
for most of the {\it Kepler} candidates. We include these FPPs in the
final column of Tables \ref{c1cand}--\ref{c4cand}. \citet{mor16}
adopts the criteria of \citet{row14} that considers candidates with a
FPP$<$1\% as validated at the 99\% level. Note that the automated
methodology of \citet{mor16} does not utilize follow-up observations
and can calculate artificially high FPPs for candidates in
multi-planet systems if dynamical interactions between the planets
cause transit timing variations. Additionally, relatively large values
of $R_p/R_\star$ result in high FPP values (see for example
Kepler-283~c in Table~\ref{c1cand} with an FPP of 100\%). Thus, a high
FPP for a confirmed planet does not mean the confirmation is
erroneous; rather it indicates that the information is insufficient to
adequately determine the candidate disposition. It should further be
noted that the \citet{mor16} FPPs specifically relate to astrophysical
false positives linked to transits and eclipses. As such, very low
signal-to-noise candidates can actually be due to instrumental noise
or stellar variability and still have low FPP values.

Shown in Figure~\ref{histograms} are the distribution of planet radii
of all the {\it Kepler} candidate planets, represented by the vertical
gray bars, compared with the {\it Kepler} candidate planets of all
sizes within the optimistic HZ or their host stars
(Table~\ref{c4cand}), represented by the vertical green bars. The bin
edges have been set with regard to the proposed standardization of
occurrence rate bins of the NASA ExoPAG Study Analysis Group 13
(private communication). The $i^{th}$ bin in planet radius is defined
as the interval $R_i = [1.5^{i-2}, 1.5^{i-1}] R_\oplus$. This implies
the following bin edges (rounded to 2 significant figures and in units
of $R_\oplus$): 0.67, 1.0, 1.5, 2.3, 3.4, 5.1, 7.6, 11, 17, 26.

We fit a power law to both distributions, represented in
Figure~\ref{histograms} as dashed lines for all {\it Kepler}
candidates and solid lines for the HZ candidates. The power law fits
excluded the first two bins due to a lack of completeness in the data
sample for planets smaller than 1.5 $R_\oplus$. We use a power law of
the form
\begin{equation}
  \frac{d N}{d \log R_p} = k R_p^{\alpha},
\end{equation}
with similar notation to that used by \citet{how12}, where $k$ and
$\alpha$ are the power law coefficients. Note that in our case $N$
represents the total number of planets in each radius bin. Thus, this
is not intended to represent completeness but rather compare directly
the radius distributions between those candidates in the HZ with those
from the general {\it Kepler} candidate sample. The power law fits are
shown in the histograms of Figure~\ref{histograms} where the dashed
line fit uses $k = 2775$ and $\alpha = -1.44$, and the solid line fit
uses $k = 57.6$ and $\alpha = -0.93$. The large difference in $k$ is
attributable to the difference in population sizes. The difference in
the slope of the power laws, $\alpha$, would imply that smaller
planets are rarer in the HZ than in the general population. However,
the transit signal of terrestrial planets in this region is much
harder to detect due to the fewer transits that occurred during the
primary mission. Additionally, the orbital periods of planets in the
HZ can often correspond with the rotation of the {\it Kepler}
spacecraft over a complete solar orbit, resulting in significant
systematic noise. To quantify the difference in the power laws for the
two distributions, we calculated the maximum-likelihood estimator
(MLE) for each sample \citep{bau07}. For the complete sample of {\it
  Kepler} candidates, we calculate a value of $\mathrm{MLE} = 0.68 \pm
0.01$. For the HZ candidates represented in Table~\ref{c4cand}, we
calculate a value of $\mathrm{MLE} = 0.75 \pm 0.08$. Based upon the
similarity of the distributions and the MLE calculations, we conclude
that there is little evidence of a significant difference in the
populations.

\begin{deluxetable*}{llccccccc}
  \tablewidth{0pc}
  \tablecaption{\label{c1cand} Category 1 HZ candidates: $R_p < 2
    R_\oplus$, conservative HZ}
  \tablehead{
    \colhead{KOI Name} &
    \colhead{Kepler Name} &
    \colhead{$P$} &
    \colhead{$a$} &
    \colhead{$R_p$} &
    \colhead{$T_\mathrm{eff}$} &
    \colhead{$R_\star$} &
    \colhead{$F_p$} &
    \colhead{FPP} \\
    \colhead{} &
    \colhead{} &
    \colhead{(days)} &
    \colhead{(AU)} &
    \colhead{($R_\oplus$)} &
    \colhead{(K)} &
    \colhead{($R_\odot$)} &
    \colhead{($F_\oplus$)} &
    \colhead{(\%)}
  }
  \startdata
K00571.05 & Kepler-186 f & $ 129.94411\pm 0.00125$ & $0.432$ & $ 1.17\pm 0.08$ & $3755\pm  90$ & $ 0.52\pm 0.02$ & $ 0.26\pm 0.04$ &  15.840 \\
K00701.04 & Kepler-62 f  & $ 267.29099\pm 0.00500$ & $0.718$ & $ 1.41\pm 0.07$ & $4925\pm  70$ & $ 0.64\pm 0.02$ & $ 0.42\pm 0.05$ &   0.122 \\
K01298.02 & Kepler-283 c & $  92.74371\pm 0.00141$ & $0.341$ & $ 1.82\pm 0.12$ & $4351\pm 100$ & $ 0.57\pm 0.02$ & $ 0.90\pm 0.15$ & 100.000 \\
K01422.04 & Kepler-296 f & $  63.33627\pm 0.00061$ & $0.255$ & $ 1.80\pm 0.31$ & $3740\pm 130$ & $ 0.48\pm 0.08$ & $ 0.62\pm 0.29$ &   0.067 \\
K02418.01 &              & $  86.82899\pm 0.00107$ & $0.290$ & $ 1.25\pm 0.21$ & $3576\pm  78$ & $ 0.46\pm 0.05$ & $ 0.37\pm 0.11$ &   0.712 \\
K02626.01 &              & $  38.09724\pm 0.00029$ & $0.158$ & $ 1.27\pm 0.45$ & $3554\pm  76$ & $ 0.40\pm 0.05$ & $ 0.91\pm 0.31$ &  27.690 \\
K03010.01 &              & $  60.86617\pm 0.00052$ & $0.247$ & $ 1.56\pm 0.17$ & $3808\pm  73$ & $ 0.52\pm 0.03$ & $ 0.84\pm 0.17$ &   0.253 \\
K03138.01 &              & $   8.68907\pm 0.00003$ & $0.038$ & $ 0.57\pm 0.04$ & $2703\pm   0$ & $ 0.12\pm 0.00$ & $ 0.47\pm 0.00$ &   2.724 \\
K03497.01 &              & $  20.35973\pm 0.00006$ & $0.129$ & $ 0.61\pm 0.13$ & $3419\pm  72$ & $ 0.34\pm 0.06$ & $ 0.87\pm 0.38$ &   0.105 \\
K04036.01 &              & $ 168.81117\pm 0.00127$ & $0.540$ & $ 1.70\pm 0.15$ & $4798\pm  95$ & $ 0.71\pm 0.03$ & $ 0.82\pm 0.14$ &   0.277 \\
K04356.01 &              & $ 174.50984\pm 0.00185$ & $0.484$ & $ 1.90\pm 0.28$ & $4367\pm 140$ & $ 0.45\pm 0.05$ & $ 0.28\pm 0.09$ &   0.315 \\
K04742.01 & Kepler-442 b & $ 112.30530\pm 0.00260$ & $0.409$ & $ 1.34\pm 0.14$ & $4402\pm 100$ & $ 0.60\pm 0.02$ & $ 0.73\pm 0.11$ &  59.110 \\
K06343.01 &              & $ 569.45154\pm 0.05848$ & $1.356$ & $ 1.90\pm 0.55$ & $6117\pm 191$ & $ 0.95\pm 0.19$ & $ 0.61\pm 0.32$ &   1.048 \\
K06425.01 &              & $ 521.10828\pm 0.04224$ & $1.217$ & $ 1.50\pm 0.44$ & $5942\pm 169$ & $ 0.95\pm 0.20$ & $ 0.68\pm 0.36$ &   0.481 \\
K06676.01 &              & $ 439.21979\pm 0.01354$ & $1.138$ & $ 1.77\pm 0.42$ & $6546\pm 178$ & $ 0.94\pm 0.17$ & $ 1.14\pm 0.53$ &  39.220 \\
K07223.01 &              & $ 317.05838\pm 0.00731$ & $0.835$ & $ 1.50\pm 0.28$ & $5366\pm 152$ & $ 0.71\pm 0.09$ & $ 0.55\pm 0.19$ &   1.802 \\
K07235.01 &              & $ 299.70688\pm 0.03513$ & $0.825$ & $ 1.15\pm 0.26$ & $5608\pm 152$ & $ 0.76\pm 0.10$ & $ 0.76\pm 0.29$ &   8.719 \\
K07470.01 &              & $ 392.50116\pm 0.03343$ & $1.002$ & $ 1.90\pm 0.93$ & $5128\pm 161$ & $ 0.99\pm 0.40$ & $ 0.60\pm 0.56$ &   3.805 \\
K07554.01 &              & $ 482.62012\pm 0.03127$ & $1.233$ & $ 1.94\pm 0.58$ & $6335\pm 197$ & $ 1.07\pm 0.23$ & $ 1.09\pm 0.61$ &   1.306 \\
K07591.01 &              & $ 328.32211\pm 0.01347$ & $0.837$ & $ 1.30\pm 0.24$ & $4902\pm 175$ & $ 0.67\pm 0.06$ & $ 0.33\pm 0.11$ &   3.146
  \enddata
\end{deluxetable*}

\begin{deluxetable*}{llccccccc}
  \tablewidth{0pc}
  \tablecaption{\label{c2cand} Category 2 HZ candidates: $R_p < 2
    R_\oplus$, optimistic HZ}
  \tablehead{
    \colhead{KOI Name} &
    \colhead{Kepler Name} &
    \colhead{$P$} &
    \colhead{$a$} &
    \colhead{$R_p$} &
    \colhead{$T_\mathrm{eff}$} &
    \colhead{$R_\star$} &
    \colhead{$F_p$} &
    \colhead{FPP} \\
    \colhead{} &
    \colhead{} &
    \colhead{(days)} &
    \colhead{(AU)} &
    \colhead{($R_\oplus$)} &
    \colhead{(K)} &
    \colhead{($R_\odot$)} &
    \colhead{($F_\oplus$)} &
    \colhead{(\%)}
  }
  \startdata
K00463.01 &              & $  18.47764\pm 0.00002$ & $0.092$ & $ 1.48\pm 0.31$ & $3395\pm  71$ & $ 0.28\pm 0.06$ & $ 1.14\pm 0.54$ &   0.005 \\
K00571.05 & Kepler-186 f & $ 129.94411\pm 0.00125$ & $0.432$ & $ 1.17\pm 0.08$ & $3755\pm  90$ & $ 0.52\pm 0.02$ & $ 0.26\pm 0.04$ &  15.840 \\
K00701.03 & Kepler-62 e  & $ 122.38740\pm 0.00080$ & $0.427$ & $ 1.61\pm 0.05$ & $4925\pm  70$ & $ 0.64\pm 0.02$ & $ 1.19\pm 0.14$ &   0.130 \\
K00701.04 & Kepler-62 f  & $ 267.29099\pm 0.00500$ & $0.718$ & $ 1.41\pm 0.07$ & $4925\pm  70$ & $ 0.64\pm 0.02$ & $ 0.42\pm 0.05$ &   0.122 \\
K01298.02 & Kepler-283 c & $  92.74371\pm 0.00141$ & $0.341$ & $ 1.82\pm 0.12$ & $4351\pm 100$ & $ 0.57\pm 0.02$ & $ 0.90\pm 0.15$ & 100.000 \\
K01422.04 & Kepler-296 f & $  63.33627\pm 0.00061$ & $0.255$ & $ 1.80\pm 0.31$ & $3740\pm 130$ & $ 0.48\pm 0.08$ & $ 0.62\pm 0.29$ &   0.067 \\
K01422.05 & Kepler-296 e & $  34.14211\pm 0.00025$ & $0.169$ & $ 1.53\pm 0.26$ & $3740\pm 130$ & $ 0.48\pm 0.08$ & $ 1.42\pm 0.67$ &  26.410 \\
K02418.01 &              & $  86.82899\pm 0.00107$ & $0.290$ & $ 1.25\pm 0.21$ & $3576\pm  78$ & $ 0.46\pm 0.05$ & $ 0.37\pm 0.11$ &   0.712 \\
K02626.01 &              & $  38.09724\pm 0.00029$ & $0.158$ & $ 1.27\pm 0.45$ & $3554\pm  76$ & $ 0.40\pm 0.05$ & $ 0.91\pm 0.31$ &  27.690 \\
K03010.01 &              & $  60.86617\pm 0.00052$ & $0.247$ & $ 1.56\pm 0.17$ & $3808\pm  73$ & $ 0.52\pm 0.03$ & $ 0.84\pm 0.17$ &   0.253 \\
K03138.01 &              & $   8.68907\pm 0.00003$ & $0.038$ & $ 0.57\pm 0.04$ & $2703\pm   0$ & $ 0.12\pm 0.00$ & $ 0.47\pm 0.00$ &   2.724 \\
K03282.01 &              & $  49.27676\pm 0.00037$ & $0.215$ & $ 1.92\pm 0.21$ & $3899\pm  78$ & $ 0.53\pm 0.03$ & $ 1.25\pm 0.26$ &   0.065 \\
K03497.01 &              & $  20.35973\pm 0.00006$ & $0.129$ & $ 0.61\pm 0.13$ & $3419\pm  72$ & $ 0.34\pm 0.06$ & $ 0.87\pm 0.38$ &   0.105 \\
K04036.01 &              & $ 168.81117\pm 0.00127$ & $0.540$ & $ 1.70\pm 0.15$ & $4798\pm  95$ & $ 0.71\pm 0.03$ & $ 0.82\pm 0.14$ &   0.277 \\
K04087.01 & Kepler-440 b & $ 101.11141\pm 0.00078$ & $0.242$ & $ 1.86\pm 0.22$ & $4134\pm 154$ & $ 0.56\pm 0.04$ & $ 1.40\pm 0.41$ &   0.422 \\
K04356.01 &              & $ 174.50984\pm 0.00185$ & $0.484$ & $ 1.90\pm 0.28$ & $4367\pm 140$ & $ 0.45\pm 0.05$ & $ 0.28\pm 0.09$ &   0.315 \\
K04427.01 &              & $ 147.66022\pm 0.00146$ & $0.419$ & $ 1.68\pm 0.21$ & $3788\pm  80$ & $ 0.49\pm 0.04$ & $ 0.25\pm 0.06$ &   2.196 \\
K04550.01 &              & $ 140.25252\pm 0.00215$ & $0.465$ & $ 1.95\pm 0.21$ & $4821\pm  81$ & $ 0.79\pm 0.04$ & $ 1.39\pm 0.22$ &   1.034 \\
K04742.01 & Kepler-442 b & $ 112.30530\pm 0.00260$ & $0.409$ & $ 1.34\pm 0.14$ & $4402\pm 100$ & $ 0.60\pm 0.02$ & $ 0.73\pm 0.11$ &  59.110 \\
K06343.01 &              & $ 569.45154\pm 0.05848$ & $1.356$ & $ 1.90\pm 0.55$ & $6117\pm 191$ & $ 0.95\pm 0.19$ & $ 0.61\pm 0.32$ &   1.048 \\
K06425.01 &              & $ 521.10828\pm 0.04224$ & $1.217$ & $ 1.50\pm 0.44$ & $5942\pm 169$ & $ 0.95\pm 0.20$ & $ 0.68\pm 0.36$ &   0.481 \\
K06676.01 &              & $ 439.21979\pm 0.01354$ & $1.138$ & $ 1.77\pm 0.42$ & $6546\pm 178$ & $ 0.94\pm 0.17$ & $ 1.14\pm 0.53$ &  39.220 \\
K07016.01 & Kepler-452 b & $ 384.84299\pm 0.00950$ & $1.046$ & $ 1.63\pm 0.22$ & $5757\pm  85$ & $ 1.11\pm 0.12$ & $ 1.11\pm 0.31$ &   0.251 \\
K07179.01 &              & $ 407.14655\pm 0.05896$ & $1.077$ & $ 1.18\pm 0.51$ & $5845\pm 185$ & $ 1.20\pm 0.30$ & $ 1.30\pm 0.81$ & 100.000 \\
K07223.01 &              & $ 317.05838\pm 0.00731$ & $0.835$ & $ 1.50\pm 0.28$ & $5366\pm 152$ & $ 0.71\pm 0.09$ & $ 0.55\pm 0.19$ &   1.802 \\
K07235.01 &              & $ 299.70688\pm 0.03513$ & $0.825$ & $ 1.15\pm 0.26$ & $5608\pm 152$ & $ 0.76\pm 0.10$ & $ 0.76\pm 0.29$ &   8.719 \\
K07470.01 &              & $ 392.50116\pm 0.03343$ & $1.002$ & $ 1.90\pm 0.93$ & $5128\pm 161$ & $ 0.99\pm 0.40$ & $ 0.60\pm 0.56$ &   3.805 \\
K07554.01 &              & $ 482.62012\pm 0.03127$ & $1.233$ & $ 1.94\pm 0.58$ & $6335\pm 197$ & $ 1.07\pm 0.23$ & $ 1.09\pm 0.61$ &   1.306 \\
K07591.01 &              & $ 328.32211\pm 0.01347$ & $0.837$ & $ 1.30\pm 0.24$ & $4902\pm 175$ & $ 0.67\pm 0.06$ & $ 0.33\pm 0.11$ &   3.146
  \enddata
\end{deluxetable*}

\begin{deluxetable*}{llccccccc}
  \tablewidth{0pc}
  \tablecaption{\label{c3cand} Category 3 HZ candidates: any radius,
    conservative HZ}
  \tablehead{
    \colhead{KOI Name} &
    \colhead{Kepler Name} &
    \colhead{$P$} &
    \colhead{$a$} &
    \colhead{$R_p$} &
    \colhead{$T_\mathrm{eff}$} &
    \colhead{$R_\star$} &
    \colhead{$F_p$} &
    \colhead{FPP} \\
    \colhead{} &
    \colhead{} &
    \colhead{(days)} &
    \colhead{(AU)} &
    \colhead{($R_\oplus$)} &
    \colhead{(K)} &
    \colhead{($R_\odot$)} &
    \colhead{($F_\oplus$)} &
    \colhead{(\%)}
  }
  \startdata
K00433.02 &              & $ 328.23996\pm 0.00036$ & $0.917$ & $11.24\pm 0.85$ & $5234\pm 103$ & $ 0.85\pm 0.06$ & $ 0.58\pm 0.13$ &   0.270 \\
K00518.03 & Kepler-174 d & $ 247.35373\pm 0.00200$ & $0.677$ & $ 2.19\pm 0.13$ & $4880\pm 126$ & $ 0.62\pm 0.03$ & $ 0.43\pm 0.09$ &   0.005 \\
K00571.05 & Kepler-186 f & $ 129.94411\pm 0.00125$ & $0.432$ & $ 1.17\pm 0.08$ & $3755\pm  90$ & $ 0.52\pm 0.02$ & $ 0.26\pm 0.04$ &  15.840 \\
K00701.04 & Kepler-62 f  & $ 267.29099\pm 0.00500$ & $0.718$ & $ 1.41\pm 0.07$ & $4925\pm  70$ & $ 0.64\pm 0.02$ & $ 0.42\pm 0.05$ &   0.122 \\
K00841.04 &              & $ 269.29425\pm 0.00423$ & $0.812$ & $ 3.09\pm 0.34$ & $5251\pm 105$ & $ 0.82\pm 0.05$ & $ 0.69\pm 0.15$ &   0.161 \\
K00854.01 &              & $  56.05605\pm 0.00025$ & $0.226$ & $ 2.05\pm 0.24$ & $3593\pm  79$ & $ 0.49\pm 0.04$ & $ 0.71\pm 0.18$ &   0.005 \\
K00868.01 &              & $ 235.99802\pm 0.00038$ & $0.613$ & $11.00\pm 0.53$ & $4245\pm  85$ & $ 0.66\pm 0.03$ & $ 0.33\pm 0.05$ &   6.834 \\
K00881.02 &              & $ 226.89047\pm 0.00110$ & $0.668$ & $ 4.68\pm 0.56$ & $5067\pm 102$ & $ 0.75\pm 0.04$ & $ 0.75\pm 0.14$ &   0.240 \\
K00902.01 &              & $  83.92508\pm 0.00014$ & $0.304$ & $ 5.04\pm 0.50$ & $3960\pm 124$ & $ 0.51\pm 0.04$ & $ 0.62\pm 0.18$ & 100.000 \\
K00959.01 &              & $  12.71379\pm 0.00000$ & $0.049$ & $ 2.31\pm 0.00$ & $2661\pm   0$ & $ 0.12\pm 0.00$ & $ 0.26\pm 0.00$ & 100.000 \\
K01126.02 &              & $ 475.95432\pm 0.02806$ & $1.037$ & $ 3.05\pm 0.61$ & $5334\pm  80$ & $ 1.00\pm 0.14$ & $ 0.68\pm 0.23$ &  92.340 \\
K01298.02 & Kepler-283 c & $  92.74371\pm 0.00141$ & $0.341$ & $ 1.82\pm 0.12$ & $4351\pm 100$ & $ 0.57\pm 0.02$ & $ 0.90\pm 0.15$ & 100.000 \\
K01422.04 & Kepler-296 f & $  63.33627\pm 0.00061$ & $0.255$ & $ 1.80\pm 0.31$ & $3740\pm 130$ & $ 0.48\pm 0.08$ & $ 0.62\pm 0.29$ &   0.067 \\
K01431.01 &              & $ 345.15988\pm 0.00041$ & $0.981$ & $ 7.77\pm 0.80$ & $5597\pm 112$ & $ 0.93\pm 0.09$ & $ 0.79\pm 0.22$ &  78.530 \\
K01466.01 &              & $ 281.56259\pm 0.00037$ & $0.752$ & $11.35\pm 0.60$ & $4810\pm  76$ & $ 0.78\pm 0.04$ & $ 0.51\pm 0.08$ &   0.017 \\
K02020.01 &              & $ 110.96546\pm 0.00115$ & $0.368$ & $ 2.12\pm 0.28$ & $4441\pm 140$ & $ 0.55\pm 0.04$ & $ 0.77\pm 0.22$ &   0.473 \\
K02078.02 &              & $ 161.51633\pm 0.00086$ & $0.496$ & $ 2.87\pm 0.22$ & $4243\pm  84$ & $ 0.64\pm 0.03$ & $ 0.48\pm 0.08$ &   0.012 \\
K02210.02 &              & $ 210.63058\pm 0.00146$ & $0.605$ & $ 3.57\pm 0.53$ & $4895\pm  78$ & $ 0.76\pm 0.04$ & $ 0.81\pm 0.13$ &   0.046 \\
K02418.01 &              & $  86.82899\pm 0.00107$ & $0.290$ & $ 1.25\pm 0.21$ & $3576\pm  78$ & $ 0.46\pm 0.05$ & $ 0.37\pm 0.11$ &   0.712 \\
K02626.01 &              & $  38.09724\pm 0.00029$ & $0.158$ & $ 1.27\pm 0.45$ & $3554\pm  76$ & $ 0.40\pm 0.05$ & $ 0.91\pm 0.31$ &  27.690 \\
K02686.01 &              & $ 211.03387\pm 0.00083$ & $0.611$ & $ 3.83\pm 0.38$ & $4658\pm  93$ & $ 0.69\pm 0.03$ & $ 0.53\pm 0.09$ &  14.822 \\
K02703.01 &              & $ 213.25766\pm 0.00105$ & $0.609$ & $ 2.85\pm 0.33$ & $4477\pm 159$ & $ 0.64\pm 0.05$ & $ 0.40\pm 0.12$ &   0.040 \\
K02762.01 &              & $ 132.99683\pm 0.00092$ & $0.452$ & $ 2.71\pm 0.58$ & $4523\pm 161$ & $ 0.66\pm 0.05$ & $ 0.80\pm 0.25$ &   0.003 \\
K02770.01 &              & $ 205.38445\pm 0.00184$ & $0.588$ & $ 2.28\pm 0.27$ & $4400\pm 157$ & $ 0.62\pm 0.05$ & $ 0.38\pm 0.12$ &   0.789 \\
K02834.01 &              & $ 136.20563\pm 0.00128$ & $0.460$ & $ 2.39\pm 0.31$ & $4648\pm 167$ & $ 0.68\pm 0.06$ & $ 0.90\pm 0.28$ &   0.169 \\
K02992.01 &              & $  82.66049\pm 0.00071$ & $0.309$ & $ 3.36\pm 0.98$ & $3952\pm  90$ & $ 0.55\pm 0.04$ & $ 0.70\pm 0.18$ &  80.400 \\
K03010.01 &              & $  60.86617\pm 0.00052$ & $0.247$ & $ 1.56\pm 0.17$ & $3808\pm  73$ & $ 0.52\pm 0.03$ & $ 0.84\pm 0.17$ &   0.253 \\
K03138.01 &              & $   8.68907\pm 0.00003$ & $0.038$ & $ 0.57\pm 0.04$ & $2703\pm   0$ & $ 0.12\pm 0.00$ & $ 0.47\pm 0.00$ &   2.724 \\
K03263.01 &              & $  76.87935\pm 0.00005$ & $0.262$ & $ 7.90\pm 1.77$ & $3638\pm  76$ & $ 0.44\pm 0.05$ & $ 0.43\pm 0.13$ &  75.070 \\
K03497.01 &              & $  20.35973\pm 0.00006$ & $0.129$ & $ 0.61\pm 0.13$ & $3419\pm  72$ & $ 0.34\pm 0.06$ & $ 0.87\pm 0.38$ &   0.105 \\
K04036.01 &              & $ 168.81117\pm 0.00127$ & $0.540$ & $ 1.70\pm 0.15$ & $4798\pm  95$ & $ 0.71\pm 0.03$ & $ 0.82\pm 0.14$ &   0.277 \\
K04356.01 &              & $ 174.50984\pm 0.00185$ & $0.484$ & $ 1.90\pm 0.28$ & $4367\pm 140$ & $ 0.45\pm 0.05$ & $ 0.28\pm 0.09$ &   0.315 \\
K04385.02 &              & $ 386.37054\pm 0.00859$ & $1.014$ & $ 3.17\pm 0.34$ & $5119\pm  82$ & $ 0.83\pm 0.05$ & $ 0.42\pm 0.08$ &   0.317 \\
K04458.01 &              & $ 358.81808\pm 0.00282$ & $0.957$ & $ 2.47\pm 0.63$ & $6056\pm 172$ & $ 0.92\pm 0.17$ & $ 1.11\pm 0.55$ &  42.920 \\
K04742.01 & Kepler-442 b & $ 112.30530\pm 0.00260$ & $0.409$ & $ 1.34\pm 0.14$ & $4402\pm 100$ & $ 0.60\pm 0.02$ & $ 0.73\pm 0.11$ &  59.110 \\
K04745.01 & Kepler-443 b & $ 177.66930\pm 0.00305$ & $0.495$ & $ 2.33\pm 0.20$ & $4723\pm 100$ & $ 0.71\pm 0.03$ & $ 0.92\pm 0.14$ &   0.155 \\
K05202.01 &              & $ 535.93726\pm 0.02765$ & $1.311$ & $ 2.52\pm 0.69$ & $5596\pm  80$ & $ 1.32\pm 0.25$ & $ 0.89\pm 0.39$ &   0.364 \\
K05236.01 &              & $ 550.85986\pm 0.00821$ & $1.355$ & $ 2.14\pm 0.36$ & $5912\pm  77$ & $ 1.12\pm 0.15$ & $ 0.74\pm 0.23$ &   4.900 \\
K05276.01 &              & $ 220.71936\pm 0.00558$ & $0.651$ & $ 2.20\pm 0.37$ & $5150\pm 184$ & $ 0.70\pm 0.08$ & $ 0.72\pm 0.26$ &   8.834 \\
K05278.01 &              & $ 281.59155\pm 0.00076$ & $0.779$ & $ 7.49\pm 1.39$ & $5330\pm 187$ & $ 0.71\pm 0.09$ & $ 0.61\pm 0.23$ &   0.995 \\
K05284.01 &              & $ 389.31119\pm 0.00206$ & $1.016$ & $ 6.42\pm 2.31$ & $5731\pm 162$ & $ 0.96\pm 0.19$ & $ 0.86\pm 0.44$ &  71.690 \\
K05416.01 &              & $  76.37804\pm 0.00183$ & $0.296$ & $ 7.22\pm 1.35$ & $3869\pm 140$ & $ 0.58\pm 0.06$ & $ 0.78\pm 0.26$ &   0.103 \\
K05622.01 &              & $ 469.63110\pm 0.01246$ & $1.112$ & $ 3.23\pm 0.75$ & $5474\pm 158$ & $ 0.76\pm 0.11$ & $ 0.38\pm 0.15$ &   0.077 \\
K05706.01 &              & $ 425.47784\pm 0.01122$ & $1.155$ & $ 3.22\pm 0.75$ & $5977\pm 201$ & $ 1.02\pm 0.19$ & $ 0.90\pm 0.46$ &   0.491 \\
K05790.01 &              & $ 178.26392\pm 0.00203$ & $0.587$ & $ 3.04\pm 0.31$ & $4899\pm  82$ & $ 0.71\pm 0.04$ & $ 0.76\pm 0.14$ &   0.618 \\
K05792.01 &              & $ 215.73711\pm 0.00137$ & $0.630$ & $ 9.67\pm 2.58$ & $4889\pm 175$ & $ 0.72\pm 0.07$ & $ 0.68\pm 0.23$ &   0.618 \\
K05850.01 &              & $ 303.22638\pm 0.00246$ & $0.878$ & $ 3.62\pm 0.64$ & $5606\pm  80$ & $ 0.95\pm 0.10$ & $ 1.03\pm 0.27$ &  43.710 \\
K05929.01 &              & $ 466.00378\pm 0.00336$ & $1.165$ & $ 5.22\pm 1.43$ & $5830\pm 158$ & $ 0.88\pm 0.16$ & $ 0.59\pm 0.27$ &  29.470 \\
K06295.01 &              & $ 204.26801\pm 0.00857$ & $0.613$ & $11.61\pm 1.49$ & $4907\pm 139$ & $ 0.73\pm 0.07$ & $ 0.73\pm 0.21$ & 100.000 \\
K06343.01 &              & $ 569.45154\pm 0.05848$ & $1.356$ & $ 1.90\pm 0.55$ & $6117\pm 191$ & $ 0.95\pm 0.19$ & $ 0.61\pm 0.32$ &   1.048 \\
K06384.01 &              & $ 566.28174\pm 0.03469$ & $1.285$ & $ 2.78\pm 0.66$ & $5830\pm 195$ & $ 0.80\pm 0.13$ & $ 0.40\pm 0.19$ &  43.340 \\
K06425.01 &              & $ 521.10828\pm 0.04224$ & $1.217$ & $ 1.50\pm 0.44$ & $5942\pm 169$ & $ 0.95\pm 0.20$ & $ 0.68\pm 0.36$ &   0.481 \\
K06676.01 &              & $ 439.21979\pm 0.01354$ & $1.138$ & $ 1.77\pm 0.42$ & $6546\pm 178$ & $ 0.94\pm 0.17$ & $ 1.14\pm 0.53$ &  39.220 \\
K06734.01 &              & $ 498.27271\pm 0.03229$ & $1.245$ & $ 2.20\pm 0.52$ & $5288\pm  79$ & $ 0.97\pm 0.10$ & $ 0.43\pm 0.11$ &   1.613 \\
K06786.01 &              & $ 455.63330\pm 0.01771$ & $1.153$ & $ 2.96\pm 0.73$ & $5883\pm 186$ & $ 0.89\pm 0.17$ & $ 0.64\pm 0.33$ &   0.413 \\
K07136.01 &              & $ 441.17368\pm 0.04754$ & $1.117$ & $ 2.83\pm 0.69$ & $5395\pm  77$ & $ 1.07\pm 0.17$ & $ 0.70\pm 0.26$ &  59.640 \\
K07223.01 &              & $ 317.05838\pm 0.00731$ & $0.835$ & $ 1.50\pm 0.28$ & $5366\pm 152$ & $ 0.71\pm 0.09$ & $ 0.55\pm 0.19$ &   1.802 \\
K07235.01 &              & $ 299.70688\pm 0.03513$ & $0.825$ & $ 1.15\pm 0.26$ & $5608\pm 152$ & $ 0.76\pm 0.10$ & $ 0.76\pm 0.29$ &   8.719 \\
K07345.01 &              & $ 377.50262\pm 0.00857$ & $1.053$ & $ 2.18\pm 0.33$ & $5751\pm  78$ & $ 0.94\pm 0.09$ & $ 0.78\pm 0.19$ &   1.365 \\
K07470.01 &              & $ 392.50116\pm 0.03343$ & $1.002$ & $ 1.90\pm 0.93$ & $5128\pm 161$ & $ 0.99\pm 0.40$ & $ 0.60\pm 0.56$ &   3.805 \\
K07554.01 &              & $ 482.62012\pm 0.03127$ & $1.233$ & $ 1.94\pm 0.58$ & $6335\pm 197$ & $ 1.07\pm 0.23$ & $ 1.09\pm 0.61$ &   1.306 \\
K07587.01 &              & $ 366.08408\pm 0.00582$ & $0.984$ & $ 2.19\pm 0.53$ & $5941\pm 198$ & $ 0.94\pm 0.20$ & $ 1.03\pm 0.57$ & 100.000 \\
K07591.01 &              & $ 328.32211\pm 0.01347$ & $0.837$ & $ 1.30\pm 0.24$ & $4902\pm 175$ & $ 0.67\pm 0.06$ & $ 0.33\pm 0.11$ &   3.146
  \enddata
\end{deluxetable*}

\begin{deluxetable*}{llccccccc}
  \tablewidth{0pc}
  \tablecaption{\label{c4cand} Category 4 HZ candidates: any radius,
    optimistic HZ}
  \tablehead{
    \colhead{KOI Name} &
    \colhead{Kepler Name} &
    \colhead{$P$} &
    \colhead{$a$} &
    \colhead{$R_p$} &
    \colhead{$T_\mathrm{eff}$} &
    \colhead{$R_\star$} &
    \colhead{$F_p$} &
    \colhead{FPP} \\
    \colhead{} &
    \colhead{} &
    \colhead{(days)} &
    \colhead{(AU)} &
    \colhead{($R_\oplus$)} &
    \colhead{(K)} &
    \colhead{($R_\odot$)} &
    \colhead{($F_\oplus$)} &
    \colhead{(\%)}
  }
  \startdata
K00087.01 & Kepler-22 b  & $ 289.86230\pm 0.00180$ & $0.849$ & $ 2.38\pm 0.13$ & $5518\pm  44$ & $ 0.98\pm 0.02$ & $ 1.11\pm 0.08$ &   2.500 \\
K00250.04 & Kepler-26 e  & $  46.82792\pm 0.00017$ & $0.220$ & $ 2.41\pm 0.15$ & $3914\pm 119$ & $ 0.51\pm 0.02$ & $ 1.13\pm 0.23$ &   0.009 \\
K00433.02 &              & $ 328.23996\pm 0.00036$ & $0.917$ & $11.24\pm 0.85$ & $5234\pm 103$ & $ 0.85\pm 0.06$ & $ 0.58\pm 0.13$ &   0.270 \\
K00463.01 &              & $  18.47764\pm 0.00002$ & $0.092$ & $ 1.48\pm 0.31$ & $3395\pm  71$ & $ 0.28\pm 0.06$ & $ 1.14\pm 0.54$ &   0.005 \\
K00518.03 & Kepler-174 d & $ 247.35373\pm 0.00200$ & $0.677$ & $ 2.19\pm 0.13$ & $4880\pm 126$ & $ 0.62\pm 0.03$ & $ 0.43\pm 0.09$ &   0.005 \\
K00571.05 & Kepler-186 f & $ 129.94411\pm 0.00125$ & $0.432$ & $ 1.17\pm 0.08$ & $3755\pm  90$ & $ 0.52\pm 0.02$ & $ 0.26\pm 0.04$ &  15.840 \\
K00683.01 &              & $ 278.12436\pm 0.00042$ & $0.851$ & $ 5.92\pm 0.97$ & $5799\pm 110$ & $ 1.05\pm 0.13$ & $ 1.55\pm 0.50$ &  73.950 \\
K00701.03 & Kepler-62 e  & $ 122.38740\pm 0.00080$ & $0.427$ & $ 1.61\pm 0.05$ & $4925\pm  70$ & $ 0.64\pm 0.02$ & $ 1.19\pm 0.14$ &   0.130 \\
K00701.04 & Kepler-62 f  & $ 267.29099\pm 0.00500$ & $0.718$ & $ 1.41\pm 0.07$ & $4925\pm  70$ & $ 0.64\pm 0.02$ & $ 0.42\pm 0.05$ &   0.122 \\
K00841.04 &              & $ 269.29425\pm 0.00423$ & $0.812$ & $ 3.09\pm 0.34$ & $5251\pm 105$ & $ 0.82\pm 0.05$ & $ 0.69\pm 0.15$ &   0.161 \\
K00854.01 &              & $  56.05605\pm 0.00025$ & $0.226$ & $ 2.05\pm 0.24$ & $3593\pm  79$ & $ 0.49\pm 0.04$ & $ 0.71\pm 0.18$ &   0.005 \\
K00868.01 &              & $ 235.99802\pm 0.00038$ & $0.613$ & $11.00\pm 0.53$ & $4245\pm  85$ & $ 0.66\pm 0.03$ & $ 0.33\pm 0.05$ &   6.834 \\
K00881.02 &              & $ 226.89047\pm 0.00110$ & $0.668$ & $ 4.68\pm 0.56$ & $5067\pm 102$ & $ 0.75\pm 0.04$ & $ 0.75\pm 0.14$ &   0.240 \\
K00902.01 &              & $  83.92508\pm 0.00014$ & $0.304$ & $ 5.04\pm 0.50$ & $3960\pm 124$ & $ 0.51\pm 0.04$ & $ 0.62\pm 0.18$ & 100.000 \\
K00959.01 &              & $  12.71379\pm 0.00000$ & $0.049$ & $ 2.31\pm 0.00$ & $2661\pm   0$ & $ 0.12\pm 0.00$ & $ 0.26\pm 0.00$ & 100.000 \\
K01126.02 &              & $ 475.95432\pm 0.02806$ & $1.037$ & $ 3.05\pm 0.61$ & $5334\pm  80$ & $ 1.00\pm 0.14$ & $ 0.68\pm 0.23$ &  92.340 \\
K01298.02 & Kepler-283 c & $  92.74371\pm 0.00141$ & $0.341$ & $ 1.82\pm 0.12$ & $4351\pm 100$ & $ 0.57\pm 0.02$ & $ 0.90\pm 0.15$ & 100.000 \\
K01411.01 &              & $ 305.07635\pm 0.00034$ & $0.913$ & $ 7.85\pm 1.30$ & $5716\pm 109$ & $ 1.15\pm 0.16$ & $ 1.53\pm 0.53$ &   8.720 \\
K01422.04 & Kepler-296 f & $  63.33627\pm 0.00061$ & $0.255$ & $ 1.80\pm 0.31$ & $3740\pm 130$ & $ 0.48\pm 0.08$ & $ 0.62\pm 0.29$ &   0.067 \\
K01422.05 & Kepler-296 e & $  34.14211\pm 0.00025$ & $0.169$ & $ 1.53\pm 0.26$ & $3740\pm 130$ & $ 0.48\pm 0.08$ & $ 1.42\pm 0.67$ &  26.410 \\
K01430.03 & Kepler-298 d & $  77.47363\pm 0.00062$ & $0.305$ & $ 2.50\pm 0.20$ & $4465\pm 100$ & $ 0.58\pm 0.03$ & $ 1.29\pm 0.25$ &   0.025 \\
K01431.01 &              & $ 345.15988\pm 0.00041$ & $0.981$ & $ 7.77\pm 0.80$ & $5597\pm 112$ & $ 0.93\pm 0.09$ & $ 0.79\pm 0.22$ &  78.530 \\
K01466.01 &              & $ 281.56259\pm 0.00037$ & $0.752$ & $11.35\pm 0.60$ & $4810\pm  76$ & $ 0.78\pm 0.04$ & $ 0.51\pm 0.08$ &   0.017 \\
K01477.01 &              & $ 169.49954\pm 0.00115$ & $0.544$ & $10.83\pm 0.95$ & $5270\pm  79$ & $ 0.79\pm 0.05$ & $ 1.45\pm 0.27$ &  12.428 \\
K01527.01 &              & $ 192.66299\pm 0.00155$ & $0.633$ & $ 2.88\pm 0.36$ & $5401\pm 107$ & $ 0.88\pm 0.08$ & $ 1.47\pm 0.37$ &   3.133 \\
K01596.02 & Kepler-309 c & $ 105.35638\pm 0.00086$ & $0.401$ & $ 2.51\pm 0.18$ & $4713\pm 100$ & $ 0.72\pm 0.04$ & $ 1.43\pm 0.28$ &   3.160 \\
K01707.02 & Kepler-315 c & $ 265.46933\pm 0.00622$ & $0.791$ & $ 4.15\pm 0.96$ & $5796\pm 108$ & $ 1.04\pm 0.20$ & $ 1.75\pm 0.80$ &   5.535 \\
K01830.02 &              & $ 198.71124\pm 0.00066$ & $0.568$ & $ 3.64\pm 0.29$ & $5180\pm 103$ & $ 0.80\pm 0.05$ & $ 1.28\pm 0.26$ &   0.042 \\
K01871.01 &              & $  92.72968\pm 0.00040$ & $0.348$ & $ 2.32\pm 0.19$ & $4580\pm  92$ & $ 0.68\pm 0.03$ & $ 1.48\pm 0.27$ &   0.018 \\
K01876.01 &              & $  82.53425\pm 0.00034$ & $0.307$ & $ 2.38\pm 0.19$ & $4316\pm  86$ & $ 0.58\pm 0.03$ & $ 1.11\pm 0.19$ &   0.072 \\
K01986.01 &              & $ 148.46034\pm 0.00069$ & $0.515$ & $ 3.54\pm 0.52$ & $5159\pm  82$ & $ 0.82\pm 0.05$ & $ 1.62\pm 0.29$ &   0.833 \\
K02020.01 &              & $ 110.96546\pm 0.00115$ & $0.368$ & $ 2.12\pm 0.28$ & $4441\pm 140$ & $ 0.55\pm 0.04$ & $ 0.77\pm 0.22$ &   0.473 \\
K02078.02 &              & $ 161.51633\pm 0.00086$ & $0.496$ & $ 2.87\pm 0.22$ & $4243\pm  84$ & $ 0.64\pm 0.03$ & $ 0.48\pm 0.08$ &   0.012 \\
K02102.01 &              & $ 187.74702\pm 0.00189$ & $0.579$ & $ 3.12\pm 0.52$ & $5303\pm 182$ & $ 0.75\pm 0.09$ & $ 1.20\pm 0.45$ &   0.042 \\
K02210.02 &              & $ 210.63058\pm 0.00146$ & $0.605$ & $ 3.57\pm 0.53$ & $4895\pm  78$ & $ 0.76\pm 0.04$ & $ 0.81\pm 0.13$ &   0.046 \\
K02418.01 &              & $  86.82899\pm 0.00107$ & $0.290$ & $ 1.25\pm 0.21$ & $3576\pm  78$ & $ 0.46\pm 0.05$ & $ 0.37\pm 0.11$ &   0.712 \\
K02626.01 &              & $  38.09724\pm 0.00029$ & $0.158$ & $ 1.27\pm 0.45$ & $3554\pm  76$ & $ 0.40\pm 0.05$ & $ 0.91\pm 0.31$ &  27.690 \\
K02686.01 &              & $ 211.03387\pm 0.00083$ & $0.611$ & $ 3.83\pm 0.38$ & $4658\pm  93$ & $ 0.69\pm 0.03$ & $ 0.53\pm 0.09$ &  14.822 \\
K02691.01 &              & $  97.44677\pm 0.00029$ & $0.373$ & $ 3.46\pm 0.73$ & $4735\pm 170$ & $ 0.69\pm 0.06$ & $ 1.53\pm 0.49$ &   6.116 \\
K02703.01 &              & $ 213.25766\pm 0.00105$ & $0.609$ & $ 2.85\pm 0.33$ & $4477\pm 159$ & $ 0.64\pm 0.05$ & $ 0.40\pm 0.12$ &   0.040 \\
K02757.01 &              & $ 234.63538\pm 0.00119$ & $0.714$ & $ 2.68\pm 0.31$ & $5422\pm 107$ & $ 0.88\pm 0.08$ & $ 1.19\pm 0.30$ &   8.112 \\
K02762.01 &              & $ 132.99683\pm 0.00092$ & $0.452$ & $ 2.71\pm 0.58$ & $4523\pm 161$ & $ 0.66\pm 0.05$ & $ 0.80\pm 0.25$ &   0.003 \\
K02770.01 &              & $ 205.38445\pm 0.00184$ & $0.588$ & $ 2.28\pm 0.27$ & $4400\pm 157$ & $ 0.62\pm 0.05$ & $ 0.38\pm 0.12$ &   0.789 \\
K02834.01 &              & $ 136.20563\pm 0.00128$ & $0.460$ & $ 2.39\pm 0.31$ & $4648\pm 167$ & $ 0.68\pm 0.06$ & $ 0.90\pm 0.28$ &   0.169 \\
K02841.01 &              & $ 159.38805\pm 0.00276$ & $0.557$ & $ 2.78\pm 0.32$ & $5135\pm  81$ & $ 0.87\pm 0.06$ & $ 1.54\pm 0.31$ &   2.286 \\
K02882.01 &              & $  75.85803\pm 0.00093$ & $0.303$ & $ 2.71\pm 0.58$ & $4474\pm 164$ & $ 0.61\pm 0.06$ & $ 1.48\pm 0.49$ &  40.990 \\
K02992.01 &              & $  82.66049\pm 0.00071$ & $0.309$ & $ 3.36\pm 0.98$ & $3952\pm  90$ & $ 0.55\pm 0.04$ & $ 0.70\pm 0.18$ &  80.400 \\
K03010.01 &              & $  60.86617\pm 0.00052$ & $0.247$ & $ 1.56\pm 0.17$ & $3808\pm  73$ & $ 0.52\pm 0.03$ & $ 0.84\pm 0.17$ &   0.253 \\
K03086.01 &              & $ 174.73210\pm 0.00277$ & $0.574$ & $ 3.23\pm 0.39$ & $5201\pm  83$ & $ 0.90\pm 0.07$ & $ 1.60\pm 0.35$ &   1.100 \\
K03138.01 &              & $   8.68907\pm 0.00003$ & $0.038$ & $ 0.57\pm 0.04$ & $2703\pm   0$ & $ 0.12\pm 0.00$ & $ 0.47\pm 0.00$ &   2.724 \\
K03263.01 &              & $  76.87935\pm 0.00005$ & $0.262$ & $ 7.90\pm 1.77$ & $3638\pm  76$ & $ 0.44\pm 0.05$ & $ 0.43\pm 0.13$ &  75.070 \\
K03282.01 &              & $  49.27676\pm 0.00037$ & $0.215$ & $ 1.92\pm 0.21$ & $3899\pm  78$ & $ 0.53\pm 0.03$ & $ 1.25\pm 0.26$ &   0.065 \\
K03497.01 &              & $  20.35973\pm 0.00006$ & $0.129$ & $ 0.61\pm 0.13$ & $3419\pm  72$ & $ 0.34\pm 0.06$ & $ 0.87\pm 0.38$ &   0.105 \\
K03663.01 & Kepler-86 b  & $ 282.52548\pm 0.00011$ & $0.845$ & $ 9.13\pm 0.93$ & $5725\pm 108$ & $ 0.91\pm 0.09$ & $ 1.12\pm 0.31$ &   0.000 \\
K03726.01 &              & $ 115.99435\pm 0.00005$ & $0.419$ & $14.69\pm 1.08$ & $4530\pm 159$ & $ 0.74\pm 0.05$ & $ 1.17\pm 0.32$ &  21.640 \\
K03823.01 &              & $ 202.11754\pm 0.00034$ & $0.667$ & $ 5.79\pm 0.62$ & $5536\pm  79$ & $ 0.92\pm 0.08$ & $ 1.59\pm 0.38$ &  33.580 \\
K04016.01 &              & $ 125.41312\pm 0.00042$ & $0.420$ & $ 2.69\pm 0.24$ & $4641\pm  79$ & $ 0.75\pm 0.03$ & $ 1.32\pm 0.20$ &   0.282 \\
K04036.01 &              & $ 168.81117\pm 0.00127$ & $0.540$ & $ 1.70\pm 0.15$ & $4798\pm  95$ & $ 0.71\pm 0.03$ & $ 0.82\pm 0.14$ &   0.277 \\
K04051.01 &              & $ 163.69235\pm 0.00138$ & $0.563$ & $ 2.87\pm 0.29$ & $4999\pm  79$ & $ 0.84\pm 0.05$ & $ 1.25\pm 0.23$ &   0.396 \\
K04054.01 &              & $ 169.13345\pm 0.00140$ & $0.569$ & $ 2.04\pm 0.19$ & $5171\pm 103$ & $ 0.80\pm 0.05$ & $ 1.27\pm 0.26$ &   2.210 \\
K04084.01 &              & $ 214.88655\pm 0.00311$ & $0.696$ & $ 3.08\pm 0.50$ & $5323\pm  79$ & $ 1.00\pm 0.12$ & $ 1.47\pm 0.45$ &   0.062 \\
K04087.01 & Kepler-440 b & $ 101.11141\pm 0.00078$ & $0.242$ & $ 1.86\pm 0.22$ & $4134\pm 154$ & $ 0.56\pm 0.04$ & $ 1.40\pm 0.41$ &   0.422 \\
K04103.01 &              & $ 184.77185\pm 0.00155$ & $0.568$ & $ 2.56\pm 0.25$ & $5273\pm 105$ & $ 0.80\pm 0.05$ & $ 1.38\pm 0.29$ &   0.608 \\
K04121.01 &              & $ 198.08878\pm 0.00246$ & $0.626$ & $ 3.47\pm 0.53$ & $5275\pm  83$ & $ 0.97\pm 0.11$ & $ 1.67\pm 0.47$ &   0.038 \\
K04356.01 &              & $ 174.50984\pm 0.00185$ & $0.484$ & $ 1.90\pm 0.28$ & $4367\pm 140$ & $ 0.45\pm 0.05$ & $ 0.28\pm 0.09$ &   0.315 \\
K04385.02 &              & $ 386.37054\pm 0.00859$ & $1.014$ & $ 3.17\pm 0.34$ & $5119\pm  82$ & $ 0.83\pm 0.05$ & $ 0.42\pm 0.08$ &   0.317 \\
K04427.01 &              & $ 147.66022\pm 0.00146$ & $0.419$ & $ 1.68\pm 0.21$ & $3788\pm  80$ & $ 0.49\pm 0.04$ & $ 0.25\pm 0.06$ &   2.196 \\
K04458.01 &              & $ 358.81808\pm 0.00282$ & $0.957$ & $ 2.47\pm 0.63$ & $6056\pm 172$ & $ 0.92\pm 0.17$ & $ 1.11\pm 0.55$ &  42.920 \\
K04550.01 &              & $ 140.25252\pm 0.00215$ & $0.465$ & $ 1.95\pm 0.21$ & $4821\pm  81$ & $ 0.79\pm 0.04$ & $ 1.39\pm 0.22$ &   1.034 \\
K04742.01 & Kepler-442 b & $ 112.30530\pm 0.00260$ & $0.409$ & $ 1.34\pm 0.14$ & $4402\pm 100$ & $ 0.60\pm 0.02$ & $ 0.73\pm 0.11$ &  59.110 \\
K04745.01 & Kepler-443 b & $ 177.66930\pm 0.00305$ & $0.495$ & $ 2.33\pm 0.20$ & $4723\pm 100$ & $ 0.71\pm 0.03$ & $ 0.92\pm 0.14$ &   0.155 \\
K05202.01 &              & $ 535.93726\pm 0.02765$ & $1.311$ & $ 2.52\pm 0.69$ & $5596\pm  80$ & $ 1.32\pm 0.25$ & $ 0.89\pm 0.39$ &   0.364 \\
K05236.01 &              & $ 550.85986\pm 0.00821$ & $1.355$ & $ 2.14\pm 0.36$ & $5912\pm  77$ & $ 1.12\pm 0.15$ & $ 0.74\pm 0.23$ &   4.900 \\
K05276.01 &              & $ 220.71936\pm 0.00558$ & $0.651$ & $ 2.20\pm 0.37$ & $5150\pm 184$ & $ 0.70\pm 0.08$ & $ 0.72\pm 0.26$ &   8.834 \\
K05278.01 &              & $ 281.59155\pm 0.00076$ & $0.779$ & $ 7.49\pm 1.39$ & $5330\pm 187$ & $ 0.71\pm 0.09$ & $ 0.61\pm 0.23$ &   0.995 \\
K05284.01 &              & $ 389.31119\pm 0.00206$ & $1.016$ & $ 6.42\pm 2.31$ & $5731\pm 162$ & $ 0.96\pm 0.19$ & $ 0.86\pm 0.44$ &  71.690 \\
K05416.01 &              & $  76.37804\pm 0.00183$ & $0.296$ & $ 7.22\pm 1.35$ & $3869\pm 140$ & $ 0.58\pm 0.06$ & $ 0.78\pm 0.26$ &   0.103 \\
K05475.01 &              & $ 448.30356\pm 0.00416$ & $1.085$ & $ 2.63\pm 0.72$ & $6072\pm 152$ & $ 1.29\pm 0.32$ & $ 1.71\pm 1.02$ &   0.715 \\
K05552.01 &              & $ 295.95807\pm 0.00202$ & $0.815$ & $ 2.15\pm 0.37$ & $5505\pm 104$ & $ 0.99\pm 0.12$ & $ 1.22\pm 0.40$ &   1.840 \\
K05581.01 &              & $ 374.87625\pm 0.00711$ & $1.053$ & $ 4.92\pm 2.01$ & $5636\pm 171$ & $ 1.35\pm 0.36$ & $ 1.50\pm 0.97$ &   0.275 \\
K05622.01 &              & $ 469.63110\pm 0.01246$ & $1.112$ & $ 3.23\pm 0.75$ & $5474\pm 158$ & $ 0.76\pm 0.11$ & $ 0.38\pm 0.15$ &   0.077 \\
K05706.01 &              & $ 425.47784\pm 0.01122$ & $1.155$ & $ 3.22\pm 0.75$ & $5977\pm 201$ & $ 1.02\pm 0.19$ & $ 0.90\pm 0.46$ &   0.491 \\
K05790.01 &              & $ 178.26392\pm 0.00203$ & $0.587$ & $ 3.04\pm 0.31$ & $4899\pm  82$ & $ 0.71\pm 0.04$ & $ 0.76\pm 0.14$ &   0.618 \\
K05792.01 &              & $ 215.73711\pm 0.00137$ & $0.630$ & $ 9.67\pm 2.58$ & $4889\pm 175$ & $ 0.72\pm 0.07$ & $ 0.68\pm 0.23$ &   0.618 \\
K05850.01 &              & $ 303.22638\pm 0.00246$ & $0.878$ & $ 3.62\pm 0.64$ & $5606\pm  80$ & $ 0.95\pm 0.10$ & $ 1.03\pm 0.27$ &  43.710 \\
K05929.01 &              & $ 466.00378\pm 0.00336$ & $1.165$ & $ 5.22\pm 1.43$ & $5830\pm 158$ & $ 0.88\pm 0.16$ & $ 0.59\pm 0.27$ &  29.470 \\
K06295.01 &              & $ 204.26801\pm 0.00857$ & $0.613$ & $11.61\pm 1.49$ & $4907\pm 139$ & $ 0.73\pm 0.07$ & $ 0.73\pm 0.21$ & 100.000 \\
K06343.01 &              & $ 569.45154\pm 0.05848$ & $1.356$ & $ 1.90\pm 0.55$ & $6117\pm 191$ & $ 0.95\pm 0.19$ & $ 0.61\pm 0.32$ &   1.048 \\
K06384.01 &              & $ 566.28174\pm 0.03469$ & $1.285$ & $ 2.78\pm 0.66$ & $5830\pm 195$ & $ 0.80\pm 0.13$ & $ 0.40\pm 0.19$ &  43.340 \\
K06425.01 &              & $ 521.10828\pm 0.04224$ & $1.217$ & $ 1.50\pm 0.44$ & $5942\pm 169$ & $ 0.95\pm 0.20$ & $ 0.68\pm 0.36$ &   0.481 \\
K06676.01 &              & $ 439.21979\pm 0.01354$ & $1.138$ & $ 1.77\pm 0.42$ & $6546\pm 178$ & $ 0.94\pm 0.17$ & $ 1.14\pm 0.53$ &  39.220 \\
K06734.01 &              & $ 498.27271\pm 0.03229$ & $1.245$ & $ 2.20\pm 0.52$ & $5288\pm  79$ & $ 0.97\pm 0.10$ & $ 0.43\pm 0.11$ &   1.613 \\
K06786.01 &              & $ 455.63330\pm 0.01771$ & $1.153$ & $ 2.96\pm 0.73$ & $5883\pm 186$ & $ 0.89\pm 0.17$ & $ 0.64\pm 0.33$ &   0.413 \\
K07016.01 & Kepler-452 b & $ 384.84299\pm 0.00950$ & $1.046$ & $ 1.63\pm 0.22$ & $5757\pm  85$ & $ 1.11\pm 0.12$ & $ 1.11\pm 0.31$ &   0.251 \\
K07040.01 &              & $ 502.20642\pm 0.04742$ & $1.152$ & $ 3.61\pm 1.44$ & $6346\pm  82$ & $ 1.21\pm 0.14$ & $ 1.62\pm 0.46$ &  60.420 \\
K07136.01 &              & $ 441.17368\pm 0.04754$ & $1.117$ & $ 2.83\pm 0.69$ & $5395\pm  77$ & $ 1.07\pm 0.17$ & $ 0.70\pm 0.26$ &  59.640 \\
K07179.01 &              & $ 407.14655\pm 0.05896$ & $1.077$ & $ 1.18\pm 0.51$ & $5845\pm 185$ & $ 1.20\pm 0.30$ & $ 1.30\pm 0.81$ & 100.000 \\
K07223.01 &              & $ 317.05838\pm 0.00731$ & $0.835$ & $ 1.50\pm 0.28$ & $5366\pm 152$ & $ 0.71\pm 0.09$ & $ 0.55\pm 0.19$ &   1.802 \\
K07235.01 &              & $ 299.70688\pm 0.03513$ & $0.825$ & $ 1.15\pm 0.26$ & $5608\pm 152$ & $ 0.76\pm 0.10$ & $ 0.76\pm 0.29$ &   8.719 \\
K07345.01 &              & $ 377.50262\pm 0.00857$ & $1.053$ & $ 2.18\pm 0.33$ & $5751\pm  78$ & $ 0.94\pm 0.09$ & $ 0.78\pm 0.19$ &   1.365 \\
K07470.01 &              & $ 392.50116\pm 0.03343$ & $1.002$ & $ 1.90\pm 0.93$ & $5128\pm 161$ & $ 0.99\pm 0.40$ & $ 0.60\pm 0.56$ &   3.805 \\
K07554.01 &              & $ 482.62012\pm 0.03127$ & $1.233$ & $ 1.94\pm 0.58$ & $6335\pm 197$ & $ 1.07\pm 0.23$ & $ 1.09\pm 0.61$ &   1.306 \\
K07587.01 &              & $ 366.08408\pm 0.00582$ & $0.984$ & $ 2.19\pm 0.53$ & $5941\pm 198$ & $ 0.94\pm 0.20$ & $ 1.03\pm 0.57$ & 100.000 \\
K07591.01 &              & $ 328.32211\pm 0.01347$ & $0.837$ & $ 1.30\pm 0.24$ & $4902\pm 175$ & $ 0.67\pm 0.06$ & $ 0.33\pm 0.11$ &   3.146
  \enddata
\end{deluxetable*}


\section{Dynamical Stability of HZ Candidates}
\label{stability}

Of the 29 HZ candidates from category 2 (radii less than $2 R_\oplus$
and within the optimistic HZ; Table~\ref{c2cand}), 6 are in
multi-planet systems. Specifically, 3 five-planet systems (Kepler-62,
Kepler-186 and Kepler-296) and 1 double system (Kepler-283c) harbor
these 6 candidates. For the candidates of any radii within the
optimistic HZ (Table~\ref{c4cand}), 19 are in multi-planet systems (13
double systems, 4 triple systems, and 2 quadruple system). Six of
these candidates from Table~\ref{c2cand} and four from
Table~\ref{c4cand} have been confirmed, however only a few have had a
thorough dynamical stability analysis performed
\citep{bol14,bol15,shi16}. Here we examine the orbital stability of
all HZ candidates that orbit in multi-planet systems. For the small
($< 2 R_\oplus$) candidates, we further explore long-term stability
for a wide range of plausible eccentricities and compositions.

To perform stability analyses, we first need to provide masses for the
planets, as transit photometry only provides planetary radii. The
candidates (at least those from Table~\ref{c2cand}) are too small to
induce gravitational perturbations on their star or on adjacent
planets, so neither radial velocity observations nor transit timing
variations can be used to constrain their masses. We therefore turn to
mass-radius relations of the form
\begin{equation}
  M_p = M_\oplus (R_p/R_\oplus)^\alpha
\end{equation}
where $M_p$ and $R_p$ are the mass and radius of the planet,
respectively, and $\alpha$ is a model-dependent exponent. We tested
for stability using several models for $\alpha$ that were derived
theoretically \citep{val06,for07,lis11} and empirically \citep{wei14}
for completeness.

For two-planet systems, the criterion for stability is that their
separations $\Delta$ exceed about 3.5 mutual Hill radii ($R_{H,M_p}$)
\citep{gla93}, where
\begin{equation}
  R_{H,M_p} = 0.5 \; (a_{in} + a_{out}) \;[(M_{p,in} + M_{p,out})/3
    M_\star]^{\frac{1}{3}}
\end{equation}
and
\begin{equation}
  \Delta = (a_{out} - a_{in}) / R_H 
\end{equation}
Here $a$ is the semimajor axis, $M_p$ is the planet mass, $M_\star$ is
the central mass and `in/out' subscripts represent the inner and outer
planets. The two-planet system from Table~\ref{c2cand} (Kepler-283c)
and all two-planet systems from Table~\ref{c4cand} obey this
constraint, with $\Delta$ values ranging from 19--116.

For systems with more than two planets, \citet{smi09} established a
heuristic criterion of $\Delta \sim 9$ between adjacent planets in
order to have long-term stability on Gyr timescales. In some cases
$\Delta$ can be lower if an adjacent $\Delta$ is higher, so they
imposed a criterion of $\Delta_{in} + \Delta_{out} > 18$ for three
adjacent planets. The five-planet systems from Table~\ref{c2cand}
(Kepler-186, Kepler 62, and Kepler-296) all satisfy these criteria, as
do all multi-planet systems from Table~\ref{c4cand}.

\subsection{Eccentricities}

For the multi-planets systems in Table~\ref{c2cand}, we numerically
explored the dynamical stability using the \textit{Mercury}
integration package \citep{cha99} in order to examine the effect of
higher eccentricities on the long-term survival of the systems. Using
masses derived from the \citet{lis11} mass-radius relation, we
explored stability for a full range of eccentricities assigned to the
HZ candidates, simulating each case with all other planets in the
system on nearly circular and coplanar orbits. Note that higher
eccentricities for the candidate will likely induce, or will be a
result of, planet-planet interactions, however our goal is to examine
the maximum eccentricity value that could destabilize the system. We
evolved each system forward in time for 10$^{10}$ orbits of the
outermost planet using a time-step of 1/20 times the orbital period of
the innermost planet. Constraints (upper limits) on eccentricities
from these simulations are 0.3 for both Kepler-62e and f, 0.62 for
Kepler-186f, 0.72 for Kepler-283c, and 0.14 and 0.16 for Kepler-296 e
and f, respectively. Note that eccentric orbits for planets within the
HZ can produce seasonal variations that inhibit the consistent
presence of liquid water on the surface \citep{wil02,kan12b,bol16}.

\subsection{Densities}

We also explored stability for a wide range of plausible compositions
for the planets with radii $< 2 R_\oplus$. By adopting a planetary
composition model, an estimate of the planet mass is obtained whilst
providing insight into possible interior structures. Data from the
dozens of exoplanets that have both measured masses and radii (and
therefore densities), combined with theoretical models, suggest that
planets with radii less than about $1.6 R_\oplus$ are likely composed
of some combination of ice, silicate rock and iron and devoid of
massive gaseous H/He envelopes \citep{rog15}. While the HZ candidates
with radii closer to $2 R_\oplus$ are likely H/He-rich sub-Neptunes,
in theory they could still be rocky, as thermal evolution models
predict a hard upper limit for the size of an envelope-free planet at
about $2 R_\oplus$ \citep{lop14}. For our stability analysis of the
candidates from Table~\ref{c2cand}, we assume the planets haven't
accreted enough gas to significantly alter their radii. Using
radius-composition curves from \citet{for07}, we explored the
stability of these systems using compositions with different ratios of
ice, rock and iron (from pure ice to pure iron). For nearly all
systems, the extreme case of pure iron planets allowed long-term
stability for all planets in the system. The exception is Kepler-296
in which the highest density for all planets that allowed long-term
stability was a 50/50\% Earth-like/iron composition. Kepler-296 is the
most compact system of the multi-planet systems from
Table~\ref{c2cand}, so stability is more sensitive to higher
densities.

Finally, we ran long term simulations of the multi-planet systems from
Table~\ref{c4cand}, assuming nominal masses from \citet{lis11} and
nearly circular and coplanar orbits. Nearly all of the candidates from
this set have sizes within 2.2--4.7~$R_\oplus$, so fall into the
super-Earth/sub-Neptune regime, with the exception of one giant planet
candidate at $11.2 R_\oplus$. All of the systems remained stable for
the duration of the simulations.


\section{Conclusions}
\label{conclusions}

The {\it Kepler} mission has provided an enormous amount of data and
discoveries that have enabled statistical studies of exoplanets in the
terrestrial regime. Although the primary mission duration of {\it
  Kepler} was not as long as desired, the duration was sufficient for
the orbital period sensitivity to reach into the HZ of the host
stars. The primary mission goal of Kepler was thus achieved and has
provideed important insights into the frequency of terrestrial planets
in the HZ of late-type stars.

Here we have provided a concise description of HZ boundaries and
provide a catalog of {\it Kepler} candidates that lie in the HZ of
their host stars. The four different categories of candidates allow
the reader to adopt the criteria that are most useful for a particular
follow-up program. For example, giant planets in the optimistic HZ
(Table~\ref{c4cand}) may be useful for those interesting in HZ
exomoons where a wider range of incident flux can account for
additional energy sources from tidal energy, etc
\citep{hel13,hin13}. Our analysis of the radii distributions for
candidates in the HZ compared with the general candidate population
shows that the two are very similar within the constraints of
selection effects and systematic noise that impacts longer-period
terrestrial planets. Our dynamical stability simulations are
consistent with all of the multi-planet systems with a planet in the
HZ being stable for reasonable assumptions regarding the planet
densities and compositions.

Recall that the HZ is primarily a target selection tool rather than
any guarantee regarding habitability. Similar catalogs, such as the
Catalog of Earth-Like Exoplanet Survey Targets (CELESTA) provided by
\citet{cha16} are intended for the design of further missions and
observing strategies that will ultimately lead to detailed exoplanet
characterization. The utility of catalogs such as the one provided
here is to inform the community of the distribution of planetary
objects that occupy the HZ and encourage further follow-up and
validation of the candidates that remain to be confirmed.


\section*{Acknowledgements}

The authors would like to thank the anonymous referee, whose comments
greatly improved the quality of the paper. The authors also thank
Douglas Caldwell and Timothy Morton for enlightening discussions
regarding {\it Kepler} candidate validation. Nader Haghighipour
acknowledges support from NASA ADAP grant NNX13AF20G. James Kasting
and Ravi Kopparapu gratefully acknowledge funding from NASA
Astrobiology Institute's Virtual Planetary Laboratory lead team,
supported by NASA under cooperative agreement NNH05ZDA001C. Ravi
Kopparapu also acknowledges support from NASA Habitable Worlds grant
NNH14ZDA001N-HW. {\it Kepler} is NASA's 10th Discovery Mission and was
funded by the agency's Science Mission Directorate. This research has
made use of the NASA Exoplanet Archive and the ExoFOP site, which are
operated by the California Institute of Technology, under contract
with the National Aeronautics and Space Administration under the
Exoplanet Exploration Program. This work has also made use of the
Habitable Zone Gallery at hzgallery.org. The results reported herein
benefited from collaborations and/or information exchange within
NASA's Nexus for Exoplanet System Science (NExSS) research
coordination network sponsored by NASA's Science Mission Directorate.


\end{document}